\documentclass[aps,pre,twocolumn,groupedaddress]{revtex4}

\usepackage{graphicx}
\usepackage{bm}
\usepackage{amsmath}
\usepackage{verbatim}
\usepackage{color}

\begin{document}

\title{Active Brownian particles near straight or curved walls: Pressure and boundary layers}
\author{Ayhan Duzgun}
\author{Jonathan V. Selinger}
\affiliation{Liquid Crystal Institute, Kent State University, Kent, Ohio 44242, USA}

\date{December 27, 2017}

\begin{abstract}
Unlike equilibrium systems, active matter is not governed by the conventional laws of thermodynamics. Through a series of analytic calculations and Langevin dynamics simulations, we explore how systems cross over from equilibrium to active behavior as the activity is increased. In particular, we calculate the profiles of density and orientational order near straight or circular walls, and show the characteristic width of the boundary layers. We find a simple relationship between the enhancements of density and pressure near a wall. Based on these results, we determine how the pressure depends on wall curvature, and hence make approximate analytic predictions for the motion of curved tracers, as well as the rectification of active particles around small openings in confined geometries.   
\end{abstract}

\maketitle

\section{Introduction}

The statistical properties of active particles are quite different from the analogous properties of passive particles~\cite{Marchetti2013,Bechinger2016}.  For example, by the conventional laws of thermodynamics, equilibrium Brownian motion cannot perform any useful work.  By contrast, active Brownian motion has been shown to power microscopic gears, thus performing mechanical work~\cite{Angelani2009,Sokolov2010,Maggi2015}.  Similarly, active Brownian particles drive the motion of curved tracers~\cite{Mallory2014}, and induce flexible membranes to fold~\cite{Mallory2015}.  Parallel plates immersed in a bath of active Brownian particles experience an attractive depletion force, analogous to the Casimir effect, unlike plates in a bath of equilibrium Brownian particles~\cite{Ray2014}.  A bath of active particles exerts an active pressure on the walls, which is not a state function of the fluid but rather depends on properties of the walls~\cite{Solon2015}, particularly on the wall curvature~\cite{Fily2014,Fily2015,Smallenburg2015,yan_brady_2015}.  Many of these phenomena arise from a distinctive feature of active systems that self-propelled particles accumulate at walls~\cite{Berke2008}.

The purpose of this article is to explore how systems cross over from equilibrium to active behavior, as a function of particle activity, through several example calculations.  First, in Secs.~II and III, we review the derivation of equations of motion for order parameter fields in a system of active particles, and use these equations to calculate the steady-state solutions in several specific geometries:  near a straight wall, between two straight walls, inside and outside a circular wall.  These calculations show the formation of boundary layers near the walls, which are characterized by enhancements in the density and polar order.  We note that similar calculations have been done previously~\cite{yan_brady_2015,ezhilan_saintillan_2015}; here we review them and verify them through simulation, so that we can apply them to calculate forces and densities in specific geometries in the following sections.  (Most of these calculations are done by truncating a series of equations for moments of the density distribution, but the Appendix shows that some results apply even without this truncation.)

In Sec.~IV, we find a simple relationship between the density enhancement in the boundary layer and the active pressure on a wall.  Through this relationship, we determine the pressure on a straight wall, as well as inside and outside a circular wall.  We combine these calculations into a single curvature-dependent active pressure.  This result is consistent with previous results by other investigators, but obtained through a different method.

In Sec.~V, we apply these findings to specific geometries that demonstrate important differences between equilibrium and active systems.  For a pair of parallel plates in a bath of active Brownian particles, we use boundary-layer considerations to estimate the Casimir-like depletion force between the plates.  For a curved tracer in an active bath, we find the net force resulting from the different pressures on the inner and outer surfaces.  Finally, for a circular particle corral with just a small opening, we show that particle activity leads to a difference in densities between inside and outside.

\section{Theoretical formalism}

We begin by reviewing the derivation of equations of motion for a system of active Brownian particles, beginning with Langevin dynamics for individual particles and leading up to equations for order parameter fields that can be solved in arbitrary geometries.

We consider a system of active, non-interacting Brownian particles in two dimensions (2D).  We suppose that each particle has a position $\bm{r}=(x,y)$ and an orientation $\hat{\bm{n}}=(\cos\theta,\sin\theta)$ for its self-propulsive force.  Apart from this self-propulsive force, the particles are isotropic.  To describe the time evolution of the position and orientation, we use Langevin stochastic dynamics in the overdamped limit, which gives
\begin{subequations}
\label{langevin}
\begin{align}
&\frac{d\bm{r}}{d t}=\beta D_t \left[F\hat{\bm{n}}-\bm{\nabla}U(\bm{r})+\bm{f}(t)\right],\\
&\frac{d\theta}{d t}=\beta D_r \left[g(t)\right].
\end{align}
\end{subequations}
Here, $D_t$ and $D_r$ are the translational and rotational diffusion constants, respectively, and $\beta=1/(k_B T)$ is the inverse temperature.  The coefficient $F$ is the self-propulsive force of each particle, and the product $v_0 =\beta D_t F$ is the self-propulsive velocity.  The function $U(\bm{r})$ is the potential energy, so that $-\bm{\nabla}U(\bm{r})$ is the force derived from the potential.  The final terms $\bm{f}(t)$ and $g(t)$ are the random force and torque acting on each particle.  These stochastic contributions satisfy Gaussian white noise statistics, such that
\begin{align}
&\langle f_i(t)\rangle=0,\quad\langle g(t)\rangle=0,\nonumber\\
&\langle f_i(t)f_j(t')\rangle=\frac{2\delta_{ij}\delta(t-t')}{\beta^2 D_t},\quad\langle g(t)g(t')\rangle=\frac{2\delta(t-t')}{\beta^2 D_r},\nonumber\\
&\langle f_i(t)g(t')\rangle=0.
\end{align}

To characterize a large ensemble of active Brownian particles, we use the probability distribution function $P(\bm{r},\theta,t)$, which gives the probability of finding a particle at position $\bm{r}$ with orientation $\theta$ at time $t$.  This distribution function evolves in time following the Smoluchowski equation
\begin{align}
\frac{\partial P}{\partial t}=&-\frac{\partial}{\partial r_i}\left[\frac{\langle\Delta r_i\rangle}{\Delta t}P\right]
-\frac{\partial}{\partial\theta}\left[\frac{\langle\Delta\theta\rangle}{\Delta t}P\right]\nonumber\\
&+\frac{1}{2}\frac{\partial^2}{\partial r_i \partial r_j}\left[\frac{\langle\Delta r_i \Delta r_j \rangle}{\Delta t}P\right]
+\frac{1}{2}\frac{\partial^2}{\partial\theta^2}\left[\frac{\left\langle\Delta\theta^2\right\rangle}{\Delta t}P\right]\nonumber\\
&+\frac{\partial^2}{\partial r_i \partial\theta}\left[\frac{\langle\Delta r_i \Delta\theta\rangle}{\Delta t}P\right].
\end{align}
Here, the quantities in angle brackets are averages calculated over all particles at position $\bm{r}$ with orientation $\theta$ during a small time interval from $t$ to $t+\Delta t$.  In our system, direct integration of the Langevin equations gives
\begin{align}
&\langle\Delta r_i\rangle=(v_0 n_i-\beta D_t\partial_i U)\Delta t,\quad\langle\Delta\theta\rangle=0,\nonumber\\
&\langle\Delta r_i\Delta r_j\rangle=2D_t \delta_{ij}\Delta t,\quad\langle\Delta\theta^2\rangle=2D_r \Delta t,\nonumber\\
&\langle\Delta r_i\Delta\theta\rangle=0.
\end{align}
With those expectation values, the Smoluchowski equation becomes
\begin{align}
\frac{\partial P}{\partial t}=&-\bm{\nabla}\cdot[(v_0\hat{\bm{n}}-\beta D_t \bm{\nabla}U)P + D_t \bm{\nabla}P]
+D_r \frac{\partial^2 P}{\partial\theta^2}\nonumber\\
=&-\bm{\nabla}\cdot\bm{J}+D_r\frac{\partial^2 P}{\partial\theta^2},
\label{smoluchowski}
\end{align}
where $\bm{J}(\bm{r},\theta,t)=(v_0\hat{\bm{n}}-\beta D_t \bm{\nabla}U)P + D_t \bm{\nabla}P$ is the current density of particles at position $\bm{r}$ with orientation $\theta$ at time $t$.

As a simplification, instead of considering the full distribution $P(\bm{r},\theta,t)$ as a function of $\theta$, we calculate orientational moments of the distribution
\begin{subequations}
\begin{align}
&\rho(\bm{r},t)=\int_0^{2\pi} P(\bm{r},\theta,t)d\theta,\\
&M_i(\bm{r},t)=\int_0^{2\pi} P(\bm{r},\theta,t)n_i d\theta,\\
&Q_{ij}(\bm{r},t)=\int_0^{2\pi} P(\bm{r},\theta,t)\left(2n_i n_j -\delta_{ij}\right)d\theta,
\end{align}
\end{subequations}
with $n_1=\cos\theta$ and $n_2=\sin\theta$.  The zero-th moment $\rho(\bm{r},t)$ is the total density of particles, integrated over all orientations, as a function of positition and time.  The higher moments (normalized by $\rho$) give the orientational order parameters as functions of position and time.  In particular, the vector $\bm{M}(\bm{r},t)/\rho(\bm{r},t)$ is the polar order parameter, and the tensor $\bm{Q}(\bm{r},t)/\rho(\bm{r},t)$ is the nematic order parameter.  By integrating over the Smoluchowski equation~(\ref{smoluchowski}), we obtain equations of motion for the moments
\begin{subequations}
\label{dipolequadrupoleeqs}
\begin{align}
\frac{\partial\rho}{\partial t}=&-\partial_i \left[v_0 M_i -\beta D_t (\partial_i U)\rho -D_t\partial_i \rho\right]\nonumber\\
=&-\partial_i J^{(0)}_i,\\
\frac{\partial M_j}{\partial t}=&-\partial_i \left[\textstyle{\frac{1}{2}}v_0(\rho\delta_{ij}+Q_{ij})-\beta D_t (\partial_i U) M_j -D_t \partial_i M_j \right]\nonumber\\
&\qquad-D_r M_j\nonumber\\
=&-\partial_i J^{(1)}_{ij}-D_r M_j.
\end{align}
\end{subequations}
Here, the moments of current density are defined as $J^{(0)}_i(\bm{r},t)=\int_0^{2\pi} J_i(\bm{r},\theta,t)d\theta$ and $J^{(1)}_{ij}(\bm{r},t)=\int_0^{2\pi} J_i(\bm{r},\theta,t)n_j d\theta$.

In principle, there is an infinite series of equations of motion for the moments, with the equation for the dipole moment $M_i$ depending on the quadrupole moment $Q_{ij}$, the equation for the quadrupole moment depending on the octupole moment, and so forth.  As an approximation, we truncate the series by assuming that the quadrupole moment $Q_{ij}=0$.  With that approximation, Eqs.~(\ref{dipolequadrupoleeqs}) provide a closed set of equations for $\rho(\bm{r},t)$ and $M_i(\bm{r},t)$, which can be solved to find the distribution of active Brownian particles in any geometry.

\section{Solution in simple geometries}

In this section, we find steady-state solutions of the Smoluchowski moment equations~(\ref{dipolequadrupoleeqs}) in several specific geometries with straight or curved walls.  We consider hard walls, so that the potential energy can be written as
\begin{equation}
\label{hardwallpotential}
U(\bm{r})=
\begin{cases}
0, & \text{outside wall}, \\
\infty, & \text{inside wall}.
\end{cases}
\end{equation}
In the free space outside the wall where $U(\bm{r})=0$, in the steady state, the Smoluchowski moment equations simplify to
\begin{subequations}
\label{smoluchowskimomentssteadystate}
\begin{align}
&0=-v_0\bm{\nabla}\cdot\bm{M}+D_t \nabla^2 \rho,\\
&0=-\textstyle{\frac{1}{2}}v_0 \bm{\nabla}\rho + D_t \nabla^2 \bm{M} - D_r \bm{M}.
\end{align}
\end{subequations}
These equations can be combined to give
\begin{equation}
\nabla^4 \rho = \left(\frac{D_r}{D_t}+\frac{v_0^2}{2D_t^2}\right)\nabla^2 \rho = \xi^{-2}\nabla^2 \rho.
\end{equation}
Hence, we see that the equation has a natural length scale $\xi$, which can be written as
\begin{equation}
\xi =\frac{\sqrt{D_t/D_r}}{\sqrt{1+v_0^2/(2 D_r D_t)}}=\frac{a}{\sqrt{1+\frac{1}{2}\text{Pe}^2}}.
\end{equation}
In the numerator, the ratio of translational and rotational diffusion constants gives the length scale $a=\sqrt{D_t/D_r}$, which is typically of the same order as the particle diameter.  The denominator is expressed in terms of the Peclet number $\text{Pe}=v_0/\sqrt{D_t D_r}$, which is a dimensionless ratio that characterizes the particle activity.

The differential equations must be solved with boundary conditions expressing the constraint that particles cannot enter the hard wall.  These boundary conditions can be written in terms of the moments of current density as
\begin{subequations}
\label{bcathardwall}
\begin{align}
&0=\hat{\bm{N}}\cdot\bm{J}^{(0)}=\hat{\bm{N}}\cdot(v_0 \bm{M}-D_t \bm{\nabla}\rho),\\
&0=\hat{\bm{N}}\cdot\bm{J}^{(1)}=\hat{\bm{N}}\cdot(\textstyle{\frac{1}{2}}v_0 \rho\bm{I}-D_t \bm{\nabla}\bm{M}),
\end{align}
\end{subequations}
where $\hat{\bm{N}}$ is the local normal to the wall.  This system of equations can be solved exactly in several cases.

\subsection{Particles near an infinite straight wall}

\begin{figure}
\includegraphics[width=3.375in]{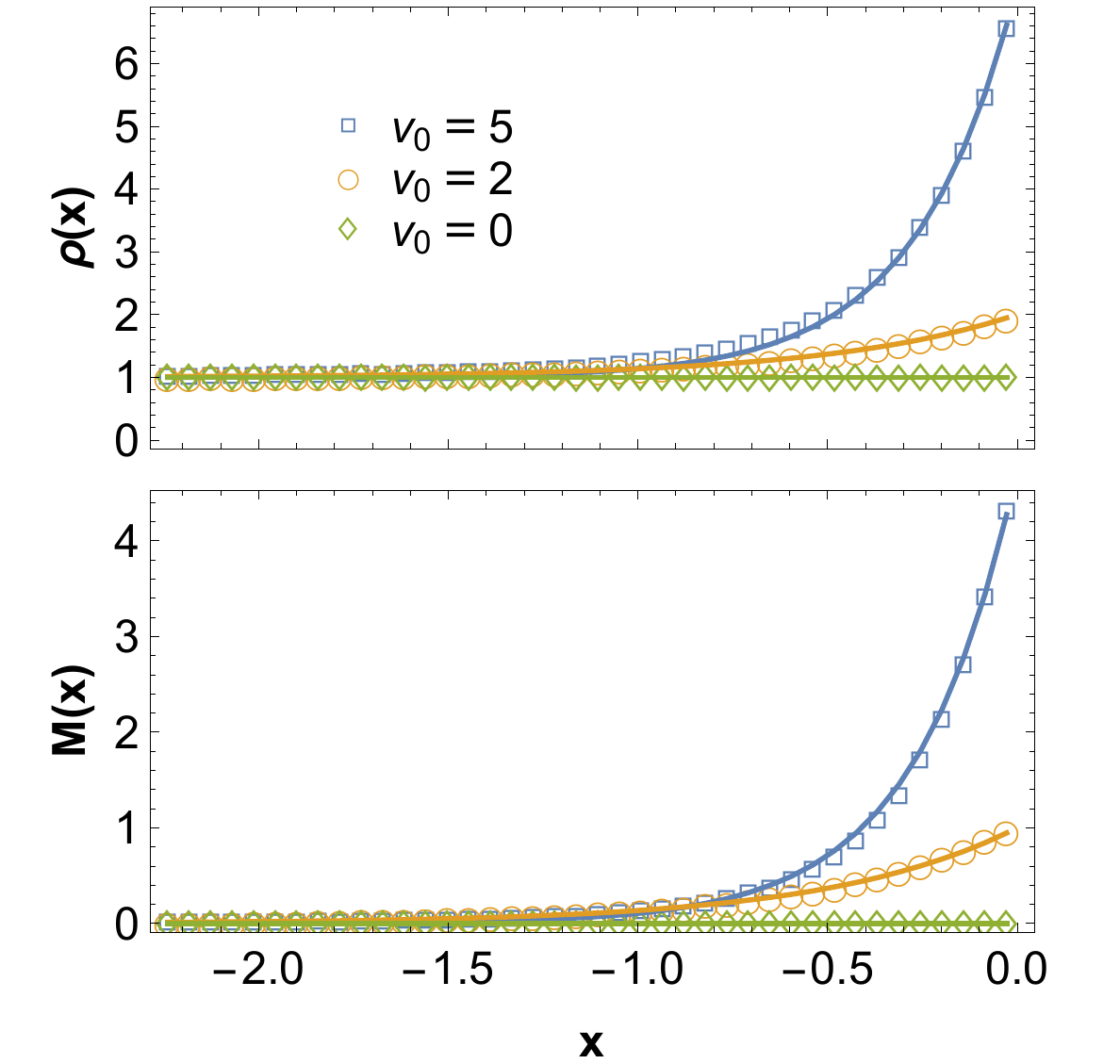}
\caption{(Color online) Plots of the density $\rho(x)$ and first moment $M_x(x)$ as functions of position $x$ near a hard wall.  Lines are the analytic predictions of Eqs.~(\ref{predictionsnearwall}), and symbols are numerical results from simulations of Langevin dynamics.  Activity is $v_0 =0$ (green diamonds), $v_0 =2$ (orange circles), and $v_0 =5$ (blue squares), and other parameters are $D_r =2$, $D_t=1$, and $\beta=1$.  All quantities are in arbitrary units.}
\label{NearWall}
\end{figure}

Consider an infinite, straight wall along the $y$-axis, so that the region $x<0$ is free space, and $x>0$ is excluded.  By the symmetry of this problem, we assume that $\rho$ and $M_x$ are functions of $x$ only, and $M_y=0$.  Using these assumptions, we solve the differential equations~(\ref{smoluchowskimomentssteadystate}) with the boundary conditions~(\ref{bcathardwall}), and the additional boundary condition that $\rho(x)\to\rho _\text{bulk}$ as $x\to -\infty$.  The solution is
\begin{subequations}
\label{predictionsnearwall}
\begin{align}
&\rho(x)=\rho_\text{bulk}\left(1+\frac{v_0^2}{2 D_r D_t}e^{x/\xi}\right),\\
&M(x)=\frac{\rho_\text{bulk}v_0}{2D_r \xi}e^{x/\xi}.
\end{align}
\end{subequations}
This solution is plotted in Fig.~\ref{NearWall} for three sample sets of parameters.  The density $\rho(x)$ is enhanced in a boundary layer of thickness $\xi$ near the wall, and decays exponentially to $\rho_\text{bulk}$.  The maximum density occurs right at the wall, where
\begin{equation}
\label{densityatwall}
\rho_\text{wall}=\rho_\text{bulk}\left(1+\frac{v_0^2}{2 D_r D_t}\right).
\end{equation}
In the Appendix, we will show that this result is exact, and does not depend on the truncation of moments.

The first moment $M_x(x)$ is nonzero in the same boundary layer, and decays exponentially to zero.  Because $M_x(x)$ is positive, we can see that particles accumulate at the wall with their orientations pointing into the wall, but are unable to enter the wall.  The orientational order parameter $M_x(x)/\rho(x)$ is greatest at the wall, and decays exponentially into the bulk, which is isotropic.

As a numerical check on the calculation, we perform simulations of the Langevin equations~(\ref{langevin}) with a boundary at $x=0$, and histograms of the density $\rho(x)$ and first moment $M_x(x)$ are plotted in Fig.~\ref{NearWall}.  The numerical results agree very well with the analytic predictions.  

\subsection{Particles between two walls}

\begin{figure}
\includegraphics[width=3.375in]{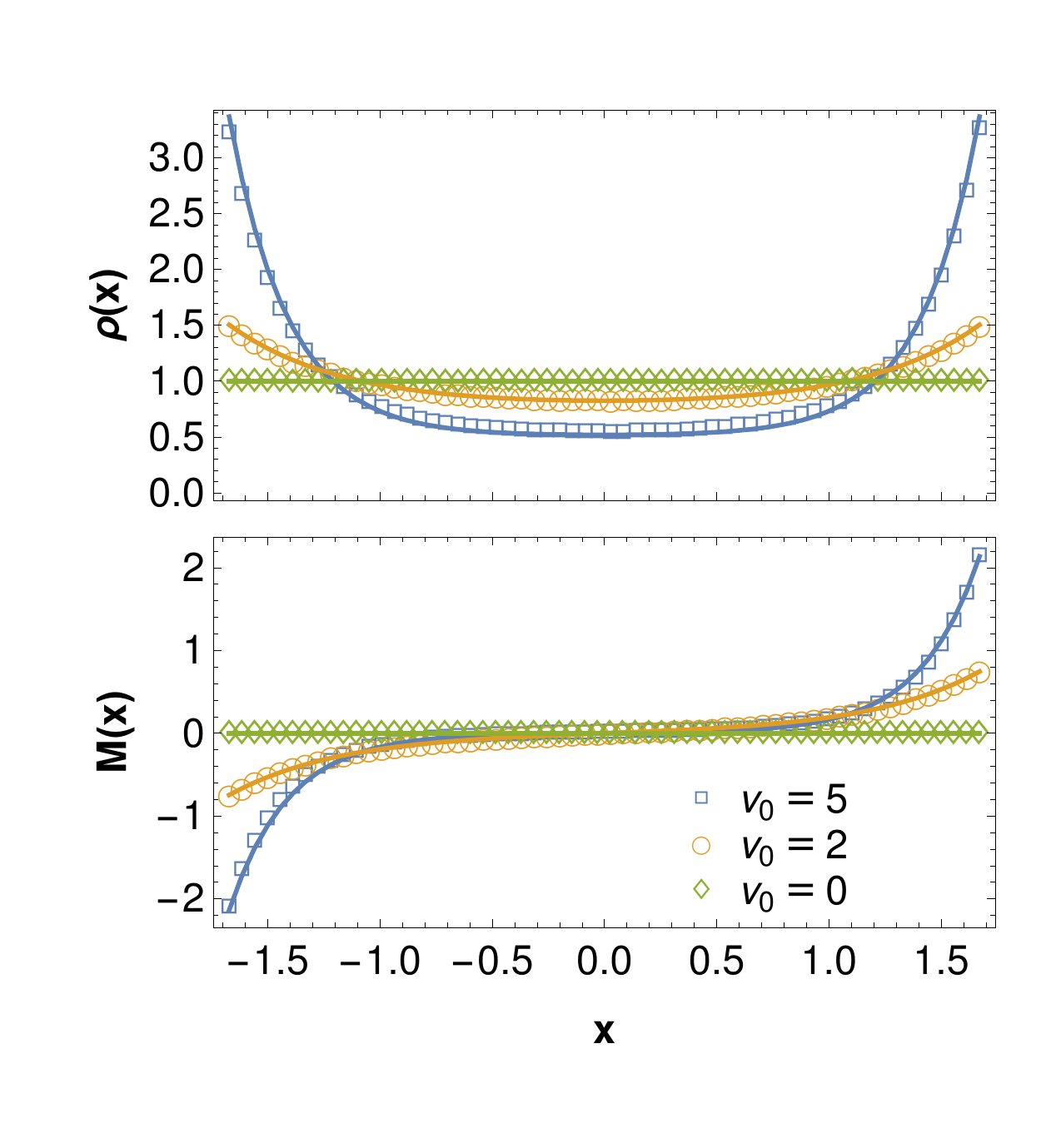}
\caption{(Color online) Plots of the density $\rho(x)$ and first moment $M_x(x)$ as functions of position $x$ between two hard walls for $\bar{\rho}=1$.  Lines are the analytic predictions of Eqs.~(\ref{predictionsbetweenwalls}--\ref{coefficientbetweenwalls}), and symbols are numerical simulation results.  Parameters are the same as in Fig.~\ref{NearWall}, and all quantities are in arbitrary units.}
\label{BetweenWalls}
\end{figure}

Now consider a system with two infinite, parallel walls at $x=\pm L$.  We solve the differential equations~(\ref{smoluchowskimomentssteadystate}) with the boundary conditions~(\ref{bcathardwall}) on both walls.  The solution is
\begin{subequations}
\label{predictionsbetweenwalls}
\begin{align}
&\rho(x)=\rho_1\left[1+\frac{v_0^2}{2 D_r D_t}\frac{\cosh(x/\xi)}{\cosh(L/\xi)}\right],\\
&M_x(x)=\frac{\rho_1 v_0}{2 D_r\xi}\frac{\sinh(x/\xi)}{\cosh(L/\xi)},
\end{align}
\end{subequations}
with an overall coefficient $\rho_1$.  This solution is plotted in Fig.~\ref{BetweenWalls}.  Here, the density $\rho(x)$ has boundary layers of thickness $\xi$ near both walls, and the first moment $M_x(x)$ points into each of the walls.

Instead of a constraint on the density far from the wall, we now have a constraint on the integrated number of particles in the system, which can be written as $\int_{-L}^L \rho(x)dx=2L\bar{\rho}$, where $\bar{\rho}$ is the average density.  Hence, the overall coefficient is
\begin{equation}
\label{coefficientbetweenwalls}
\rho_1 = \bar{\rho}\left[1+\frac{v_0^2 \xi}{2 D_r D_t L}\tanh\left(\frac{L}{\xi}\right)\right]^{-1}.
\end{equation}
For large wall separation, with $L\gg\xi[1+v_0^2 /(2 D_r D_t)]$, the density in the center is approximately independent of the walls, and we can just write the overall coefficient as $\rho_1=\rho_\text{bulk}$.  However, for smaller wall separation, the density in the center is depleted because the density on the walls is enhanced, as shown in the figure.

We perform numerical simulations of the Langevin equations with boundaries on both side of the domain.  In the simulation algorithm, when a particle crosses a boundary, we place it back in its previous location and update the orientation through the usual rotational diffusion.  The numerical results, plotted in Fig.~\ref{BetweenWalls}, agree very well with the analytic predictions.

\subsection{Particles inside circle}

Suppose that active particles are confined inside a hard, circular wall of radius $R$.  In this case, it is most convenient to work in terms of polar coordinates $(r,\theta)$.  By rotational symmetry, we expect that $\rho$ and $M_r$ are functions of $r$ only, and $M_\theta =0$.  We then express the differential equations~(\ref{smoluchowskimomentssteadystate}) in terms of polar coordinates.  A general solution for $\rho(r)$ is a linear combination of the modified Bessel functions $I_0(r/\xi)$ and $K_0(r/\xi)$, and the corresponding solution for $M_r(r)$ is a linear combination of $I_1(r/\xi)$ and $K_1(r/\xi)$.  Because the density must be finite at $r=0$, $\rho(r)$ can have only the $I_0(r/\xi)$ function, and hence $M_r(r)$ can have only $I_1(r/\xi)$.  By putting these functions into the boundary conditions~(\ref{bcathardwall}), expressed in polar coordinates, we obtain
\begin{subequations}
\label{solutioninside}
\begin{align}
&\rho(r)=\rho_2\left[1+\frac{v_0^2}{D_r D_t}\frac{I_0 (r/\xi)}{A_2}\right],\\
&M_r(r)=\frac{\rho_2 v_0}{D_r \xi}\frac{I_1 (r/\xi)}{A_2},
\end{align}
with the denominator
\begin{equation}
A_2=\left(1+\frac{v_0^2}{2D_r D_t}\right)I_2\left(\frac{R}{\xi}\right)+\left(1-\frac{v_0^2}{2D_r D_t}\right)I_0\left(\frac{R}{\xi}\right),
\end{equation}
\end{subequations}
and with an overall coefficient $\rho_2$.  This solution for density $\rho(r)$ shows a boundary layer of thickness $\xi$ inside the circular wall.  The first moment $M_x(x)$ points outward from the center, into the circular wall.

As in the previous case, the overall coefficient $\rho_2$ is fixed by the constraint on the integrated number of particles in the system, $\int_0^R2\pi r \rho(r) dr=\pi R^2 \bar{\rho}$, which gives
\begin{equation}
\rho_2 = \bar{\rho}\left[1+\frac{2 v_0^2 \xi I_1 (R/\xi)}{D_r D_t R A_2}\right]^{-1}.
\end{equation}
For a large circle, with $R\gg\xi[1+v_0^2 /(2 D_r D_t)]$, the density in the center is approximately independent of the wall, and we can write the overall coefficient as $\rho_2=\rho_\text{bulk}$.  However, for smaller radius, the density in the center is depleted because the density at the wall is enhanced.

\subsection{Particles outside circle}

As a modification of the previous case, consider active particles that are confined outside a hard, circular wall of radius $R$.  Again, we work in polar coordinates and express the solution in terms of modified Bessel functions.  Now the density must be finite, with $\rho(r)\to\rho_\text{bulk}$, as $r\to\infty$.  Hence, $\rho(r)$ can include only the $K_0(r/\xi)$ function, and $M_r(r)$ can include only $K_1(r/\xi)$.  From the boundary conditions on the circular wall, the solution becomes
\begin{subequations}
\label{solutionoutside}
\begin{align}
&\rho(r)=\rho_\text{bulk}\left[1+\frac{v_0^2}{D_r D_t} \frac{K_0(r/\xi)}{A_3}\right],\\
&M_r(r)=-\frac{\rho_\text{bulk} v_0}{D_r \xi}\frac{K_1(r/\xi)}{A_3},
\end{align}
with the denominator
\begin{equation}
A_3=\left(1+\frac{v_0^2}{2D_r D_t}\right)K_2\left(\frac{R}{\xi}\right)+\left(1-\frac{v_0^2}{2D_r D_t}\right)K_0\left(\frac{R}{\xi}\right).
\end{equation}
\end{subequations}
In this solution, the density $\rho(r)$ has a boundary layer of thickness $\xi$ outside the circular wall.  The first moment $M_x(x)$ points in toward the center, into the circular wall.

\section{Pressure on straight or curved walls}

In the previous section, we calculated the enhancement of density in boundary layers along hard walls of different shapes:  straight, inside a circle, and outside a circle.  In addition to the density enhancement, we would also like to calculate the pressure of active particles against each of these walls.  To calculate the pressure, we use a method based on the theory of Solon \emph{et al.}~\cite{Solon2015}, and consider a hard wall to be the limiting case of a soft wall.

As a first step, consider an infinite straight wall along the $y$-axis.  Instead of the hard wall potential energy of Eq.~(\ref{hardwallpotential}), we use the soft potential
\begin{equation}
U(x)=
\begin{cases}
0, & \text{for } x<0, \\
U' x, & \text{for } x>0,
\end{cases}
\end{equation}
where $U'$ is a finite positive constant.  The limit of $U'\to\infty$ will then represent a hard wall.  This assumption is similar to Ref.~\cite{Solon2015}, except that we use a linear potential while they used a quadratic potential.

We now solve the steady-state differential equations~(\ref{dipolequadrupoleeqs}) separately in the regions $x<0$ and $x>0$.  In each region, we look for solutions where the density $\rho(x)$ and the first moment $M_x(x)$ vary as $e^{\alpha x}$.  The differential equations then give a characteristic equation for $\alpha$.  In the region $x<0$, the characteristic equation is
\begin{equation}
\alpha^3-\left(\frac{D_r}{D_t}+\frac{v_0^2}{2D_t^2}\right)\alpha=0,
\end{equation}
and the solutions are $\alpha=0$ or $\pm1/\xi$.  In the region $x>0$, the characteristic equation is 
\begin{equation}
\alpha^3+2\beta U' \alpha^2-\left(\frac{D_r}{D_t}+\frac{v_0^2}{2D_t^2}-\beta^2 U'^2\right)\alpha-\frac{D_r \beta U'}{D_t}=0.
\end{equation}
For large $U'$, the solutions are
\begin{equation}
\alpha=\frac{D_r}{\beta U' D_t}+O\left(U'^{-2}\right)\text{ or }-\beta U' \pm\frac{v_0}{\sqrt{2}D_t}+O\left(U'^{-1}\right).
\end{equation}
Because $\rho$ cannot diverge as $x\to\pm\infty$, we must eliminate the negative value of $\alpha$ for $x<0$, and the positive value of $\alpha$ for $x>0$.  We then have four exponential modes with coefficients to be determined from the boundary conditions.  At the boundary $x=0$, we require that the density $\rho(x)$, the first moment $M_x(x)$, and the current moments $J^{(0)}_{x}(x)$ and $J^{(1)}_{xx}(x)$ must all be continuous (keeping in mind that the definition of current moments in Eqs.~(\ref{dipolequadrupoleeqs}) includes $U'$ terms in the region $x>0$).  We also require that $\rho(x)\to\rho_\text{bulk}$ as $x\to-\infty$, far from the wall.  Applying these boundary conditions, and assuming that $\beta U' D_t /v_0 \gg 1$, we obtain
\begin{equation}
\label{densitynearwall}
\rho(x)=
\begin{cases}
\rho_\text{bulk}\left(1+\frac{v_0^2}{2D_r D_t}e^{x/\xi}\right), & \text{for } x<0, \\
\rho_\text{bulk}\left(1+\frac{v_0^2}{2D_r D_t}\right)e^{-\beta U' x}, & \text{for } x>0.
\end{cases}
\end{equation}
The result for $x<0$ is exactly the same as previously calculated for a hard wall in Eq.~(\ref{predictionsnearwall}a).  The result for $x>0$ shows how the density decreases inside the wall, dominated by the Boltzmann distribution for large $U'$.

From these results, we can calculate the pressure of the particles on the wall.  As noted in Ref.~\cite{Solon2015}, the force of the particles on the wall is equal and opposite to the force of the wall on the particles.  Hence, the pressure can be calculated as
\begin{align}
\label{pressureeq}
p&=\int_0^\infty \rho(x)\frac{\partial U(x)}{\partial x}dx=U'\int_0^\infty\rho(x)dx\\
&=\frac{\rho_\text{bulk}}{\beta}\left(1+\frac{v_0^2}{2 D_r D_t}\right)
=\rho_\text{bulk} k_B T \left(1+\frac{v_0^2}{2 D_r D_t}\right). \nonumber
\end{align}
This result is consistent with Ref.~\cite{Solon2015}.  In Eq.~(\ref{pressureeq}), the first term is the pressure of an ideal gas without activity, $p_\text{ideal}=\rho_\text{bulk} k_B T$.  The second term is an enhancement due to the active velocity $v_0$.  Hence, the active pressure is enhanced over the ideal gas pressure by a factor of $(1+v_0^2/(2D_r D_t))$.  By comparison, Eq.~(\ref{densitynearwall}) shows that the density at the wall is enhanced over the bulk density by the same factor $\rho_\text{wall}=\rho(0)=\rho_\text{bulk}(1+v_0^2/(2D_r D_t))$.  Hence, the active pressure is simply related to the enhanced density at the wall by $p=\rho_\text{wall} k_B T$.

This relationship between pressure and enhanced density at the wall is quite general.  If we only assume that the wall potential $U(x)$ is large, diverging as $x\to\infty$, so that the density inside the wall is dominated by the Boltzmann distribution $\rho(x)=\rho_\text{wall} e^{-\beta U(x)}$, then the pressure becomes
\begin{align}
p&=\int_0^\infty \rho(x)\frac{\partial U(x)}{\partial x}dx=\rho_\text{wall} \int_0^\infty e^{-\beta U(x)}\frac{\partial U(x)}{\partial x}dx\nonumber\\
&=\rho_\text{wall} \int_0^{U(\infty)} e^{-\beta U}dU = \frac{\rho_\text{wall}}{\beta}=\rho_\text{wall} k_B T.
\end{align}

Now we apply the same considerations to active particles inside or outside a circular wall.  For particles inside a circle, the density profile $\rho(r)$ is given by Eq.~(\ref{solutioninside}a), with the denominator $A_2$ defined in Eq.~(\ref{solutioninside}c).  Suppose the radius $R$ is large enough that the coefficient $\rho_2$ can be approximated by $\rho_\text{bulk}$.  The density $\rho_\text{wall}$ is just the value of $\rho(r)$ at $r=R$, and hence the pressure on the wall is
\begin{equation}
p=\rho_\text{bulk}k_B T\left[1+\frac{\frac{v_0^2}{D_r D_t}}{1-\frac{v_0^2}{2D_r D_t}
+\left(1+\frac{v_0^2}{2D_r D_t}\right)\frac{I_2 (R/\xi)}{I_0 (R/\xi)}}\right].
\end{equation}
For $R\gg\xi$, we can use the asymptotic expansion $I_2(x)/I_0(x)\approx 1-2x^{-1}+x^{-2}$ to obtain
\begin{equation}
\label{pressureinside}
p=\rho_\text{bulk}k_B T\left[1+\frac{v_0^2}{2 D_r D_t}+\frac{v_0^2}{2 D_r^2 \xi R}+\frac{v_0^4 + D_r D_t v_0^2}{4 D_r^3 D_t R^2}\right].
\end{equation}
Hence, the pressure of active particles \emph{inside} a circular wall is \emph{increased}, compared with the active pressure on a straight wall, by an amount proportional to $1/R$.

For particles outside a circular wall, the density profile $\rho(r)$ is given by Eq.~(\ref{solutionoutside}a), with the denominator $A_3$ defined in Eq.~(\ref{solutionoutside}c).  The density $\rho_\text{wall}$ is just the value of $\rho(r)$ at $r=R$, and hence the pressure on the wall is
\begin{equation}
p=\rho_\text{bulk}k_B T\left[1+\frac{\frac{v_0^2}{D_r D_t}}{1-\frac{v_0^2}{2D_r D_t}
+\left(1+\frac{v_0^2}{2D_r D_t}\right)\frac{K_2 (R/\xi)}{K_0 (R/\xi)}}\right].
\end{equation}
For $R\gg\xi$, we can use the asymptotic expansion $K_2(x)/K_0(x)\approx 1+2x^{-1}+x^{-2}$ to obtain
\begin{equation}
\label{pressureoutside}
p=\rho_\text{bulk}k_B T\left[1+\frac{v_0^2}{2 D_r D_t}-\frac{v_0^2}{2 D_r^2 \xi R}+\frac{v_0^4 + D_r D_t v_0^2}{4 D_r^3 D_t R^2}\right].
\end{equation}
Hence, the pressure of active particles \emph{outside} a circular wall is \emph{reduced}, compared with the active pressure on a straight wall, by an amount proportional to $1/R$.

The three cases of straight wall, inside circle, and outside circle can all be combined into the single concept of a curvature-dependent active pressure.  Let us define the curvature $\kappa$ as $\kappa=0$ for a straight wall, $\kappa=+1/R$ inside a circle, and $\kappa=-1/R$ outside a circle.  The three equations (\ref{pressureeq}), (\ref{pressureinside}), and (\ref{pressureoutside}) can then be combined into the equation
\begin{align}
\label{pressurefunctionofcurvature}
p(\kappa)=\rho_\text{bulk}k_B T\biggl[&\left(1+\frac{v_0^2}{2 D_r D_t}\right)+\left(\frac{v_0^2}{2 D_r^2 \xi}\right)\kappa\nonumber\\
&+\left(\frac{v_0^4 + D_r D_t v_0^2}{4 D_r^3 D_t}\right)\kappa^2\biggr].
\end{align}
This expression can be written more compactly in terms of the Peclet number $\text{Pe}=v_0/\sqrt{D_t D_r}$ and the particle length scale $a=\sqrt{D_t/D_r}$ as
\begin{align}
p(\kappa)=\rho_\text{bulk}k_B T\Bigl[&\left(1+\textstyle{\frac{1}{2}}\text{Pe}^2\right)
+\textstyle{\frac{1}{2}}\text{Pe}^2 \sqrt{1+\textstyle{\frac{1}{2}}\text{Pe}^2} a\kappa\nonumber\\
&+\textstyle{\frac{1}{4}}\text{Pe}^2(1+\text{Pe}^2)a^2 \kappa^2\Bigr].
\end{align}
From this result, we can see that the pressure of an active fluid on a wall depends on the shape of the wall through the curvature $\kappa$.  Of course, this is not the case for an equilibrium fluid; the pressure of an equilibrium fluid depends only on fluid properties, not on the shape of the wall.

This result for the curvature-dependent active pressure is similar to a previous theoretical result of Fily \emph{et al}.\ \cite{Fily2014,Fily2015}.  They investigated the active pressure inside a convex box with variable positive curvature, and found that the local pressure is proportional to the local curvature of the boundary.  Here we see that the result also applies in a region of negative curvature, such as the outside of a circular wall.

\section{Applications}

In this section we discuss three examples, where we can make predictions for the behavior of active systems based on the concepts of pressure and boundary layers.

\subsection{Depletion force between two plates}

\begin{figure}
\includegraphics[width=3.375in]{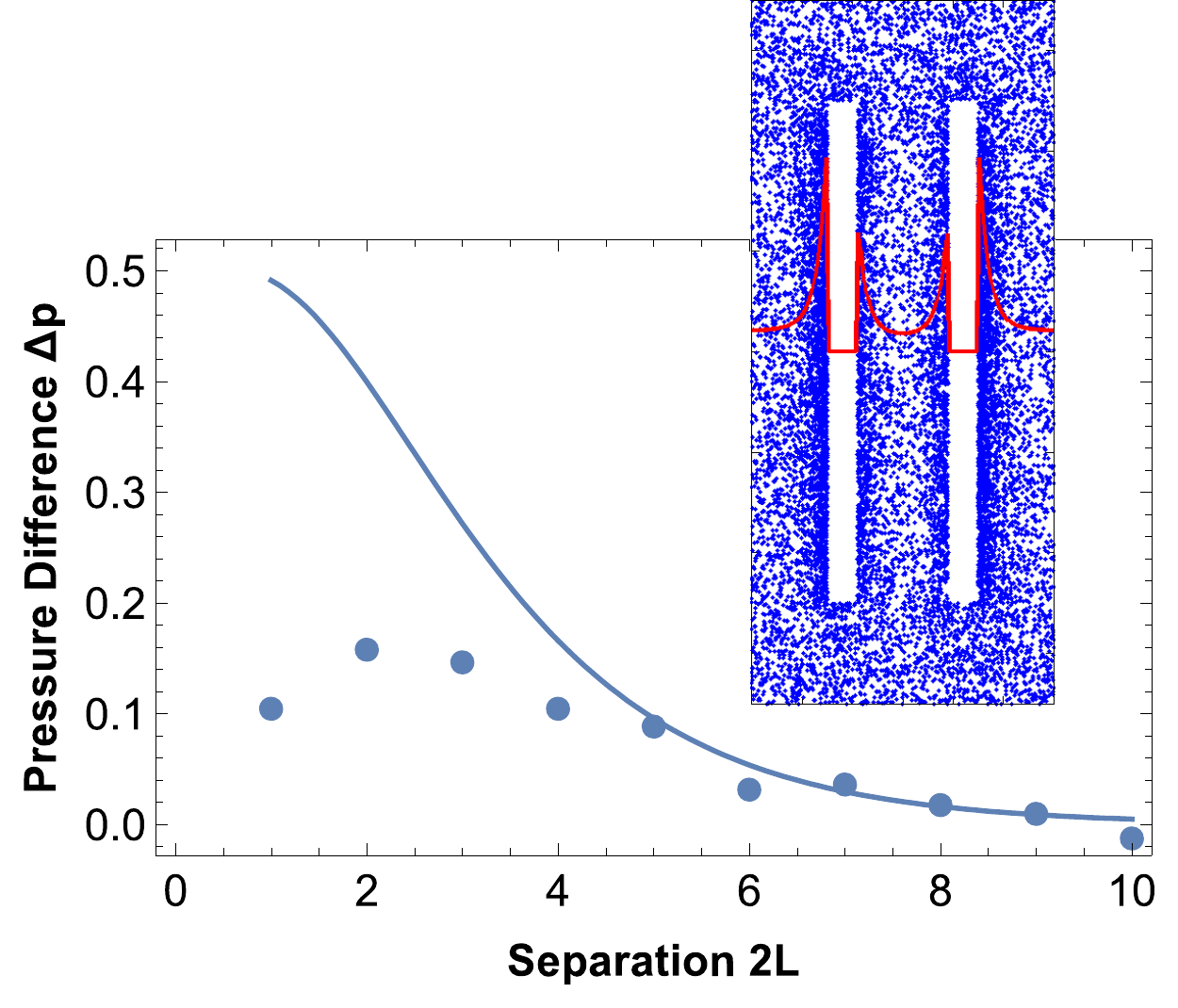}
\caption{(Color online) Theory and simulation of two parallel plates in a bath of active Brownian particles.  The inset shows a snapshot of the simulation, and the red line represents the density as a function of $x$ along the central axis $y=0$.  The main figure shows the theoretical prediction for the pressure difference as a function of the plate separation $2L$, in comparison with simulation results for the difference in densities at the inner and outer walls.  Parameters are $D_r=1$, $D_t=1$, $v_0=1$, $k_B T=1$, and plate thickness $s=0.6$.}
\label{DepletionForce}
\end{figure}

Consider two parallel plates inside a bath of active particles, separated by a distance of $2L$, as shown in Fig.~\ref{DepletionForce}.  This problem has been investigated through simulations in Ref.~\cite{Ray2014}.  Outside the plates, far from the ends, this problem is equivalent to the infinite straight wall discussed in Sec.~III(A), and the density profile is given by Eq.~(\ref{predictionsnearwall}a).  Between the plates, far from the ends, this problem is equivalent to the infinite parallel walls discussed in Sec.~III(B), and the density profile is given by Eq.~(\ref{predictionsbetweenwalls}a).  Hence, there is a boundary layer with enhanced density on each side of each plate, and thus there is active pressure on each side of each plate.  The question is:  Do all the boundary layers have the same density enhancement?  If the inner boundary layers have a different density enhancement than the outer boundary layers, then the active pressure will either push the plates together or push them apart.

To answer that question, we note that the density profiles~(\ref{predictionsnearwall}a) and~(\ref{predictionsbetweenwalls}a) have two different overall coefficients.  For the density profile outside the plates, the coefficient is $\rho_\text{bulk}$, which is the density far from the plates.  For the density profile between the plates, the coefficient is $\rho_1$.  To determine $\rho_1$, we assume that the region midway between the plates is in contact with the bulk region through the openings at the ends, so that the density midway between the plates is equal to $\rho_\text{bulk}$.  This assumption implies that the density profile between the walls is
\begin{equation}
\rho_\text{in}(x)=\frac{\rho_\text{bulk}}{1+\frac{v_0^2}{2 D_r D_t}\text{sech}(L/\xi)}\left[1+\frac{v_0^2}{2 D_r D_t}\frac{\cosh(x/\xi)}{\cosh(L/\xi)}\right].
\end{equation}
Hence, the active pressure on the inside surface of each wall is
\begin{equation}
p_\text{in}=\rho_\text{in}^\text{wall}k_B T
=\frac{\rho_\text{bulk} k_B T}{1+\frac{v_0^2}{2 D_r D_t}\text{sech}(L/\xi)}\left(1+\frac{v_0^2}{2 D_r D_t}\right).
\end{equation}
By comparison, the active pressure on the outside surface of each wall is
\begin{equation}
p_\text{out}=\rho_\text{out}^\text{wall}k_B T
=\rho_\text{bulk} k_B T \left(1+\frac{v_0^2}{2 D_r D_t}\right).
\end{equation}
Thus, the inside pressure is less than the outside pressure by
\begin{equation}
\label{casimir}
\Delta p(L)=p_\text{out}-p_\text{in}
=\frac{\rho_\text{bulk} k_B T \left(\frac{v_0^2}{2 D_r D_t}\right)\left(1+\frac{v_0^2}{2 D_r D_t}\right)}{\cosh(L/\xi)+\frac{v_0^2}{2 D_r D_t}}.
\end{equation} 
This pressure difference pushes the plates together.  It decays exponentially with $L$, with the characteristic length scale $\xi$.  Physically, this force on the plates can be regarded as a depletion force, associated with the reduced boundary layer between the plates compared with outside the plates.  It appears analogous to the Casimir force between conducting plates, but arises from a different mechanism.

We perform Langevin dynamics simulations of a bath of active Brownian particles around two parallel plates, illustrated in Fig.~\ref{DepletionForce}.  These simulations show that boundary layers form on both sides of both plates, and the outer boundary layers have higher density than the inner boundary layers, as indicated by the red line in the inset.  The relative density of these two boundary layers depends on the separation $2L$ between the plates.  The main figure presents the density difference, which is proportional to the pressure difference, in comparison with the prediction of Eq.~(\ref{casimir}).  We can see that the trends are consistent for large separation.  For smaller separation, the prediction overestimates the density difference, perhaps because it is more difficult for the density midway between the plates to become equal with $\rho_\text{bulk}$ when the openings between the plates are so small.

\subsection{Force on a curved tracer particle}

\begin{figure}
\includegraphics[width=3.375in]{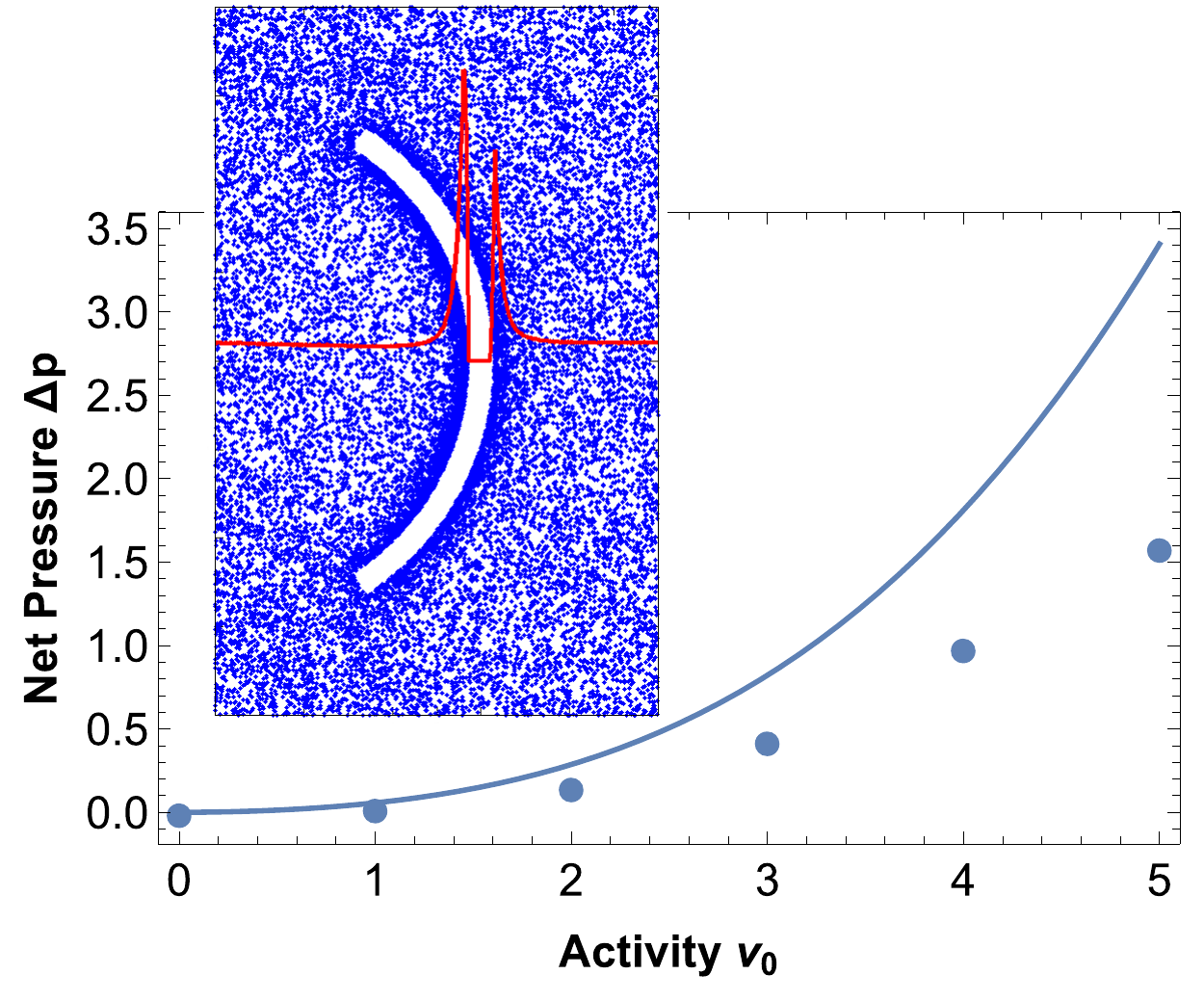}
\caption{(Color online) Theory and simulation of a curved tracer in a bath of active Brownian particles.  The inset shows the simulation, with the red line representing the density as a function of $x$ along the symmetry axis $y=0$.  The main figure shows the prediction for net pressure as a function of activity $v_0$, in comparison with the simulation results for the density difference between the two sides of the tracer.  Parameters are $D_r=2$, $D_t=1$, $k_B T=1$, tracer radius $R=7$, and tracer thickness $s=0.6$.}
\label{Tracer}
\end{figure}

Consider a curved tracer surrounded by a bath of active particles, as shown in Fig.~\ref{Tracer}.  This type of geometry has been studied through simulations in Refs.~\cite{Mallory2014,Mallory2015}.  A boundary layer forms on both sides of the tracer, and it experiences active pressure on both sides.  Based on the argument in Sec.~IV, the pressure in the inner side is greater than the pressure on the outer side.  As a result, the bath of active particles exerts a net force on the tracer, causing it to move.

To estimate the net force, we use Eq.~(\ref{pressurefunctionofcurvature}) for the pressure as a function of curvature.  On the inner side, we have the curvature $\kappa=1/(R-s/2)$, where $R$ is the radius of the midline and $s$ is the thickness of the tracer.  On the outer side, we have $\kappa=-1/(R-s/2)$.  We assume that $\rho_\text{bulk}$ is the same on both sides, because the bulk regions can easily exchange particles.  Hence, for large $R$ and small $s$, the net pressure becomes
\begin{equation}
\label{tracerprediction}
\Delta p=\frac{\rho_\text{bulk}k_B T v_0^2}{D_r^2 \xi R}\left(1+\frac{s^2}{R^2}\right).
\end{equation}

Through Langevin dynamics simulations, we visualize the distribution of active particles around the tracer and calculate the density along the symmetry axis, as indicated by the red line in the Fig.~\ref{Tracer} inset.  This simulation result shows the higher density on the inner side than on the outer side, and hence a higher pressure.  In the main figure, we show the density difference in the simulation in comparison with the prediction from Eq.~(\ref{tracerprediction}).  This results show a consistent trend, although the prediction is higher by about a factor of 2.  Hence, the approximate argument about a curvature-dependent pressure provides a simple way to understand the net pressure on the tracer.

Experiments have demonstrated that the active motion of swimming bacteria causes an asymmetric gear to rotate~\cite{Angelani2009,Sokolov2010,Maggi2015}.  The structure of the asymmetric gear is equivalent to several curved tracer particles linked together, and hence we expect that the argument in this section would also apply to the gear.

\subsection{Corral}

\begin{figure}
\includegraphics[width=3.375in]{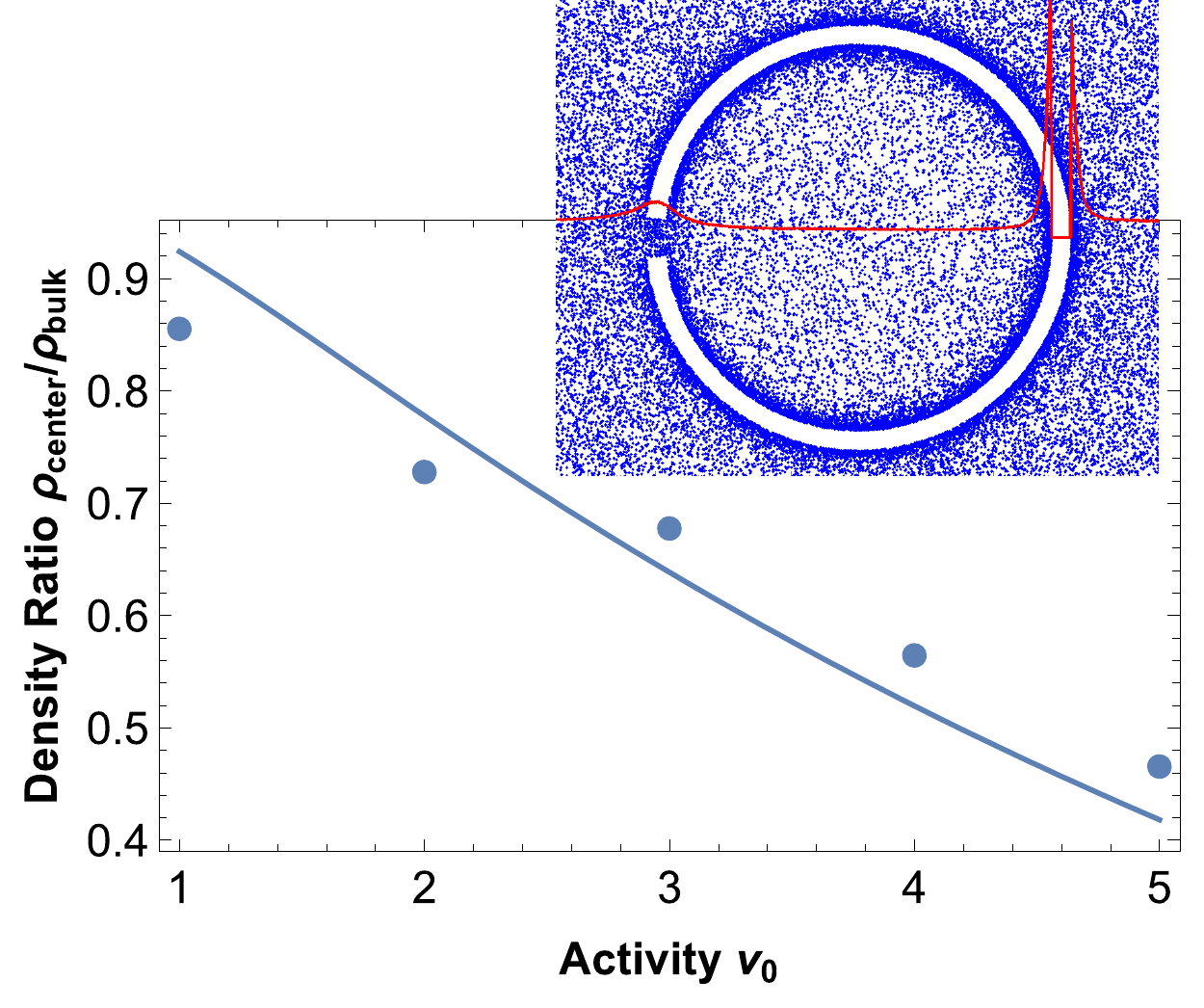}
\caption{(Color online) Theory and simulation of an active particle corral.  A snapshot of the simulation is shown in the inset, with the red line representing the density as a function of $x$ in a slice across the corral.  The main figure shows the density ratio $\rho_\text{center}/\rho_\text{bulk}$ as a function of activity $v_0$, with the points representing simulation results and the line representing the theory of Eq.~(\ref{corralprediction}).  Parameters are $D_r=2$, $D_t=1$, $k_B T=1$, and corral radius $R=4$.}
\label{Corral}
\end{figure}

As a variation on the curved tracer, consider the active particle corral shown in Fig.~\ref{Corral}.  Here, the wall is almost a full circle, with only a small opening connecting inside and outside.  The density at the inner boundary is related the the density at the center of the  circle by Eq.~(\ref{solutioninside}a), and the density at the outer boundary is related to the bulk density by Eq.~(\ref{solutionoutside}a).  The question is then:  How are the inside and outside densities related to each other?  For this geometry with only a small opening, it seems reasonable that the density of the inside boundary layer should match the density of the outside boundary layer, $\rho_\text{in}(R)=\rho_\text{out}(R)$.  This equation gives us a relationship between the density at the center of the circle and the bulk density.  For $R\gg\xi$, that relationship becomes
\begin{equation}
\label{corralprediction}
\frac{\rho_\text{center}}{\rho_\text{bulk}}=1-\frac{v_0^2}{D_r D_t}\frac{\xi}{R}.
\end{equation}
This equation shows that the corral has a reduced density at the center, compared with the bulk density.  Intuitively, this behavior occurs because it is easier for active particles to escape from the corral than to enter it, because of the shape of the opening.  It would not occur for an equilibrium fluid, which would have equal density inside and outside the corral.  Thus, the corral geometry provides one potential mechanism for tunable rectification of active particles, analogous to other mechanisms suggested by Ref.~\cite{Reichhardt_rachets}.

We carry out Langevin dynamics simulations of a bath of active particles around the corral in Fig.~\ref{Corral}.  These simulations show that the average density around the center is reduced compared with the bulk exterior, as shown by the red line.  The main figure presents the simulation results for the density ratio as a function of activity, in comparison with the prediction from Eq.~(\ref{corralprediction}), and these results are generally consistent.  Hence, the approximate argument about boundary layers provides a way to understand the relative densities in this geometry.

\acknowledgments

We would like to thank A.~Baskaran for helpful discussions.  This work was supported by National Science Foundation Grant No.~DMR-1409658. 

\appendix*
\section{}

The purpose of this appendix is to show that the relationship between the wall density $\rho_\text{wall}$ and the bulk density $\rho_\text{bulk}$ is exact in one dimension, and does not depend on the truncation of moments.

We begin with the moment equations~(\ref{dipolequadrupoleeqs}) in steady state,
\begin{subequations}
\label{currentinsteadystate}
\begin{align}
0=&-\frac{\partial J^{(0)}_x}{\partial x},\\
0=&-\frac{\partial J^{(1)}_{xx}}{\partial x}-D_r M_x .
\end{align}
\end{subequations}
Integrating Eq.~(\ref{currentinsteadystate}a) gives $J^{(0)}_x =$ constant.  Because no particles are entering or leaving the system at $x\to\pm\infty$, we must have $J^{(0)}_x = 0$.  From the definition of $J^{(0)}_x$ in Eq.~(\ref{dipolequadrupoleeqs}a), this equation implies
\begin{equation}
v_0 M_x - D_t \frac{\partial\rho}{\partial x} = 0.
\end{equation}
Integrating once again, we obtain
\begin{equation}
\label{integralone}
v_0 \int_0^\infty M_x (x) dx = D_t [\rho(\infty)-\rho(0)].
\end{equation}
By comparison, integrating Eq.~(\ref{currentinsteadystate}b) implies
\begin{equation}
\label{integraltwo}
D_r \int_0^\infty M_x (x) dx= J^{(1)}_{xx}(0)-J^{(1)}_{xx}(\infty).
\end{equation}
At the wall $x=0$, there is no current, so that $J^{(1)}_{xx}(0)=0$.  Away from the wall, the fluid becomes isotropic, and hence all the moments of the distribution function vanish, except for $\rho(\infty)=\rho_{bulk}$.  From the definition of $J^{(1)}_{xx}$ in Eq.~(\ref{dipolequadrupoleeqs}b), we obtain $J^{(1)}_{xx}(\infty)=\frac{1}{2}v_0\rho_\text{bulk}$.  Combining Eqs.~(\ref{integralone}) and~(\ref{integraltwo}) then yields
\begin{align}
&\frac{D_t}{v_0}\left[\rho_\text{bulk}-\rho(0)\right]=-\frac{v_0}{2D_r}\rho_\text{bulk},\nonumber\\
&\rho_\text{wall}=\rho(0)=\rho_\text{bulk}\left(1+\frac{v_0^2}{2 D_r D_t}\right).
\end{align}
Hence, the active pressure directly follows as
\begin{equation}
p=\rho_\text{wall}k_B T=\rho_\text{bulk}k_B T\left(1+\frac{v_0^2}{2 D_r D_t}\right).
\end{equation}

\bibliography{active3}

\end{document}